\documentclass{jpconf}
\usepackage{graphicx}
\usepackage{amsmath}
\usepackage{amssymb}
\usepackage{lmodern}

\begin{document}
\title{Matrix multiplication and universal scalability of the time on the Intel Scalable processors}
\author{Alexander Russkov$^{1}$, Lev Shchur$^{2,3}$}
\address{$^1$ Science Center in Chernogolovka, 142432 Chernogolovka, Russia} 
\address{$^2$ National Research University Higher School of Economics, 101000 Moscow, Russia}
\address{$^3$ Landau Institute for Theoretical Physics, 142432 Chernogolovka, Russia}

\begin{abstract}
Matrix multiplication is one of the core operations in many areas of scientific computing. 
We present the results of the experiments with the matrix multiplication of the big size comparable with the big size of the onboard memory, which is 1.5 terabyte in our case.
We run experiments on the computing board with two sockets and with two Intel Xeon Platinum 8164 processors, each with 26 cores and with multi-threading. The most interesting result of our study is the observation of the perfect scalability law of the matrix multiplication, and of the universality of this law.
\end{abstract}

\section{Introduction} 

At the fall of 2017 the new architecture of the computing nodes from Intel, combining the Intel(R) Xeon Scalable~\cite{Intel-X-S} architecture of the processor boards and Intel(R) Omni-Path Architecture~\cite{Intel-OPA} of the interconnections comes to the open market. It is a common thought, that these architectures will be one of the main basis for the high-performance computing in the nearest years. The new architecture becomes subject for the computer experiments with the goal to understand it's practical features, and some experiments are already reported. Application of the combination of Intel-OPA with previous architecture Intel(R) Xeon Phi discussed in~\cite{OMNI}. The experiments with the High-Performance LINPACK  (HPL) on the cluster equipped with Intel(R) Xeon Platinum 8160 processors demonstrate some interesting features of the memory organization, which shows, in particular, the increased DRAM traffic and increased L2 cache miss rates~\cite{HPL}. The technical details on the performance of the particular processors from the Intel Scalable series were published in the review~\cite{Technical}.

In this communication,  we present preliminary results of the experiments with the multiplication of the big matrices which requires a huge RAM storage of hundreds of gigabytes (GB). The main question is how the DRAM capacity boundaries and the number of cores in the processor influence the computing time.

Experiments have been done on a single node, the rack S2600WF with two Intel(R) Xeon Platinum 8164 2.0 GHz processors with 2$\times$768 GB 2666MHz DDR4 onboard memory. The software packages used are CentOS Linux release 7.5.2804, gcc compiler release 4.7.1, Intel C++ with libraries release 2018.3. 

We use Intel(R) Math Kernel Libraries (MKL)~\cite{MKL} which automatically supports features of the Intel Scalable(R) architecture, in particular, the AVX-512 extension, multi-threading, and NUMA.

\section{Number of cores influence}

\begin{figure}[htb]
	\begin{center}
\includegraphics[width=0.5\columnwidth]{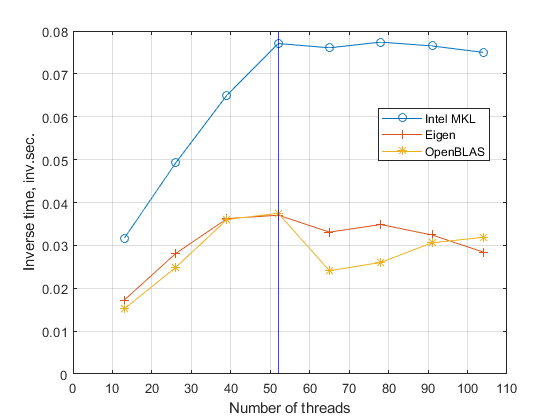}~\includegraphics[width=0.5\columnwidth]{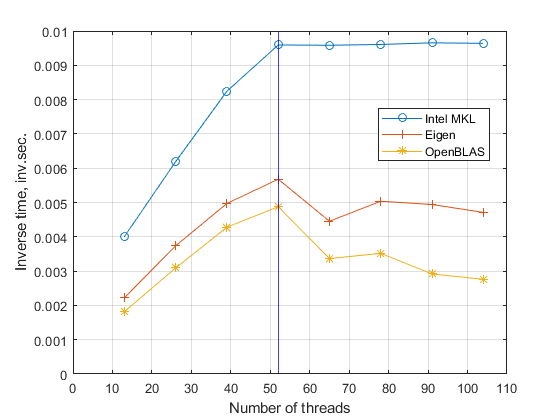}
\caption{Inverse time for the square matrix multiplication with the linear sizes 32000 (left panel) and 64000 (right panel) as function of the number of threads, for three libraries - Intel MKL~\cite{MKL}, Eigen~\cite{Eigen} and OpenBLAS~\cite{BLAS}.}
\label{fig1}
	\end{center}
\end{figure}

The Xeon Platinum 8164 processor has 26 cores and supports 52 threads. Figure~\ref{fig1} demonstrates the boundary of 52 with the multiplication of the two matrices with the linear size 32000 with the single floating point precision and of the linear size 64000 with the double floating point precision. There are two facts on the processor performance which can be extracted from the figure. Firstly, the performance is saturated for the number of threads larger than 52, clearly showing the available number of threads per one processor.  Secondly, performance is limited mainly by the memory exchange, as it is factor 8 slower at the peak performance for the matrix size 64000 compared with the matrix size 32000. This factor is due to the 8 times larger memory used for the matrix multiplication of linear size 64000, factor 4 comes from the number of matrix elements and factor two from the number representation, {\tt real*8} instead of {\tt real*4} for the linear size 32000.  We will provide more clear evidence for this observation in the following analysis. 

As seen from both panel in Fig.~\ref{fig1} the efficiency growth almost linear with the number of threads up to 52, and demonstrate almost perfect scalability.

We plot also results of matrix multiplication using Eigen library~\cite{Eigen} and OpenBLAS library~\cite{BLAS}, which demonstrates that these libraries are not using all new features of Scalable architecture yet: firstly, the scalability of performance is not linear, the efficiency grows slower than in the case of MKL, and it is saturated early for both libraries, Eigen and OpenBLAS, and even goes a bit lower for the number of threads larger than 52.

\section{The RAM size influence}

\begin{figure}[htb]
	\begin{center}
\includegraphics[width=0.5\columnwidth]{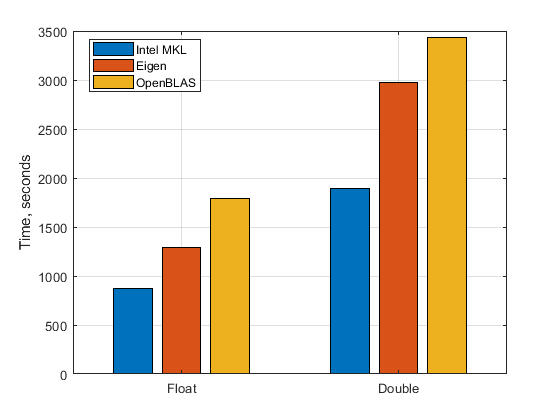}
\caption{Comparison of the times for the multiplication of the two square matrices of the linear size 128000 using the single precision (left group) and the double precision (right group) for three libraries - Intel MKL~\cite{MKL}, Eigen~\cite{Eigen} and OpenBLAS~\cite{BLAS}. The number of threads is 52.}
\label{fig2}
	\end{center}
\end{figure}

Figure~\ref{fig2} shows the time of multiplication of two square matrices of the linear size 128000. The memory necessary for matrix multiplication with the data type {\tt real*4} of elements is about 196 billion bytes (three matrices of the size 65 billion bytes), and it is 393$\cdot 10^9$ bytes for the {\tt real*8} elements. The tasks are fit within the memory accessible by a single processor, which is 768GB, accordingly Figure~\ref{fig2} does not show any influence of the memory size, and the ratio of the corresponding bars of the same color is about factor of 2, which is just the ratio of the number of bytes for two data types. At the same time, it is visible, that multiplication with the Eigen library becomes faster than with the OpenBLAS library for the larger system sizes, the trend which was already visible in the right panel of Figure~\ref{fig1}.

\begin{figure}[htb]
	\begin{center}
\includegraphics[width=0.5\columnwidth]{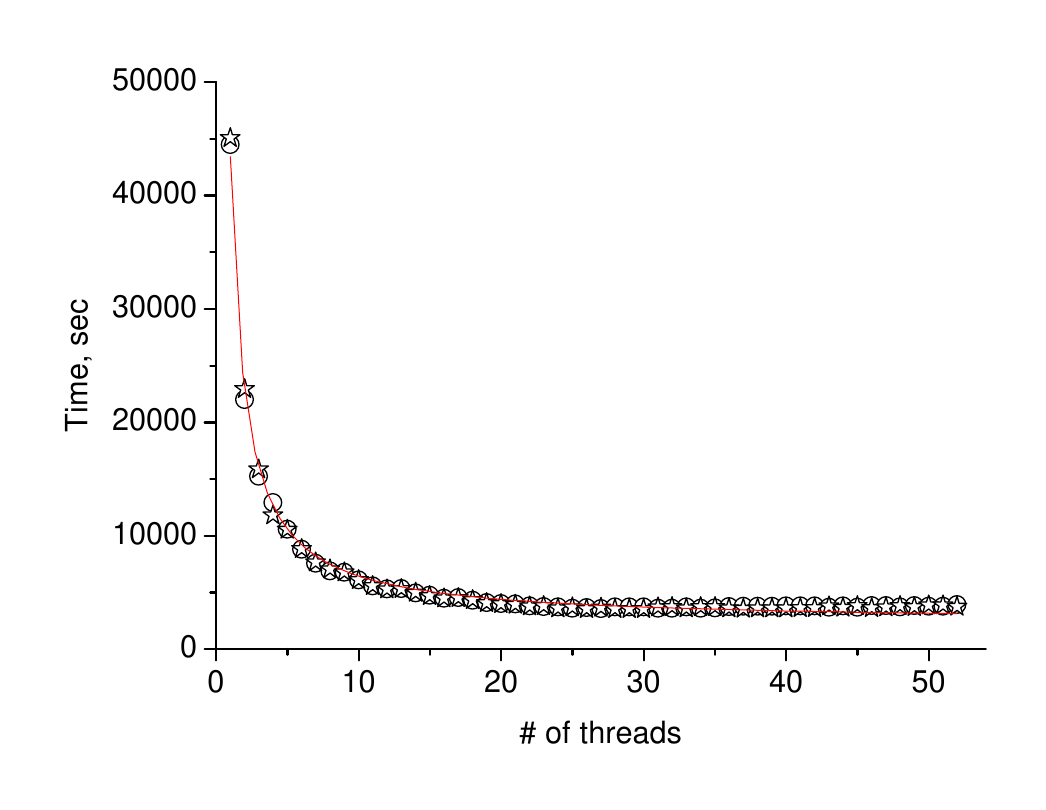}~\includegraphics[width=0.5\columnwidth]{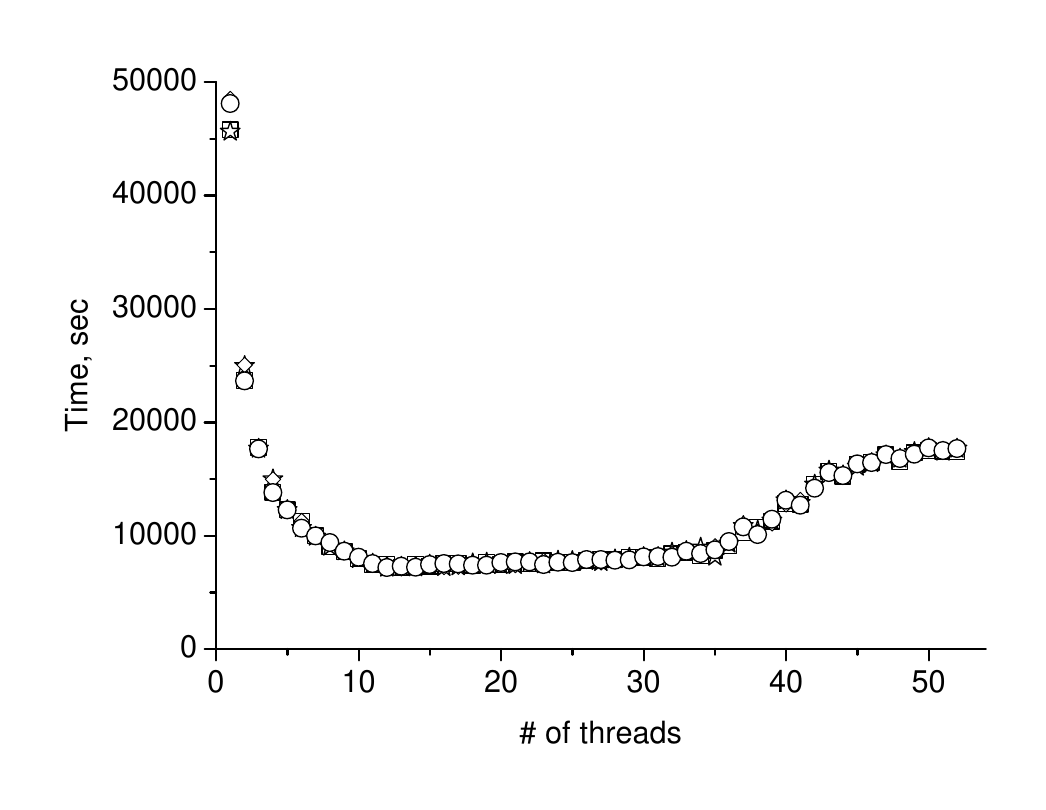}
\caption{Left panel: comparison of the times for the multiplication of the two pairs (circles and stars) of the square matrices of the linear size 128000 using the double precision. The line is the fit to the data (see the text).  Right panel: comparison of the times for the multiplication of the four pairs (circles, stars, rhombs, and squares) of the square matrices of the linear size 128000 using the double precision.}
\label{fig3}
	\end{center}
\end{figure}

In the next set of the experiments, we check how the DRAM size accessible for the single processor does influent the efficiency of the simulations. For that reason, in the first experiment, we perform in parallel two sets of matrix multiplication. We choose the size of the matrices in a way, they just fill the whole RAM. It is 6 square matrices of the linear size 128000 with {\tt real*8} elements. In total it is 786 billion bytes which just fits in the DRAM accessible by the single processor. The left panel of the Figure~\ref{fig3} shows the time variation of the two parallel processes as a function of the number of threads. Data for the time can be fit with the expression

\begin{equation}
T=T_\infty + A/N_t,
\label{expr1}
\end{equation}
i.e. inversely proportional to the number of threads $N_t$. The resulting fit is shown in the left panel of Figure~\ref{fig3} as the solid line, and fitting parameters are $T_\infty\approx 2335$ and $A\approx 41100$. The time of the multiplication achieves the asymptotic value already at the cores threshold of 26. 

Next experiment has been performed with the four sets of the same matrix multiplication running in parallel, therefore, the total memory for the matrices exceeds two times the memory directly reachable by the single processor and still fits the onboard memory of 1.5TB. The right panel of the Figure~\ref{fig3} shows the running time of the four processes (marked by circles, stars, rhombs, and squares) as function of the number of threads, and it is seen that first saturation is at the number which is half of the number of cores per processor, and the second saturation is double of the number of cores per processor, which is the maximum number of threads per processor. Probably, the saturation at the value of 13 threads is due to the memory effect but it is not clear. We plan to conduct more experiments to clear up the effect. 

\begin{figure}[htb]
	\begin{center}
\includegraphics[width=0.5\columnwidth]{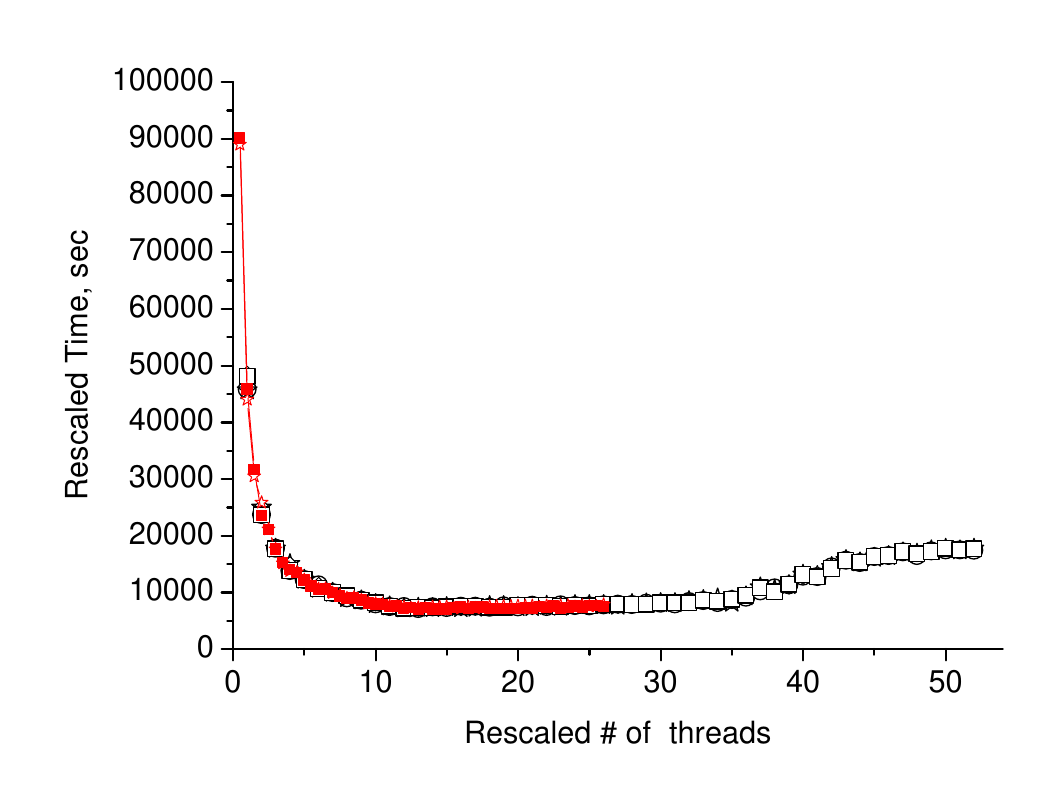}
\caption{The data collapse as the combination of the data sets from the left and right panels of the Fig~\ref{fig3}. See text for explanation.}
\label{fig4}
	\end{center}
\end{figure}

It is interesting that data in both panels of the Figure~\ref{fig3} can be collapsed on one curve. For that reason, we divide the number of threads at the left panel by factor 2 and multiply the time at the left panel by factor 2. Resulting data are shown in the Figure~\ref{fig4} by solid stars and squares, while keeping the data of the right panel of Figure~\ref{fig3} unchanged. Figure~\ref{fig4} shows excellent coincidence of the data sets. One can say that the expression~(\ref{expr1}) is the universal function for the scalability of the matrix multiplication on the Intel Scalable architecture, probably with some restrictions which are the subject of the future work.

\section{Discussion}

Presented experiments with the matrix multiplication of the big size on the Intel Xeon Platinum 8164 with 26 cores and 52 threads (the total onboard number of cores 52 and threads 104) and the big memory of 768GB per processor (total onboard memory is 1.5TB) ) shows interesting features of the Intel Scalable Architecture. We found the clear saturation of the efficiency with the number of threads larger than maximum value per processor, which is 52 in our case. We found that libraries Eigen~\cite{Eigen} and OpenBLAS~\cite{BLAS} is less effective at the moment on the Intel Scalable Architecture comparing with the Intel MKL. 

The most interesting results of the communication are the perfect scalability of the matrix multiplication time (see expression~(\ref{expr1})) and the universal behavior of the scalability as shown in the Figure~\ref{fig4}.

Future experiments can be done to understand better the performance of the Intel Scalable Architecture in computing using the matrix multiplication which is one of the cores of any scientific simulations.

\ack
The research is done within the framework of the research plan 0236-2014-0002.


\end{document}